\documentclass[aps,prl,reprint,twocolumn,superscriptaddress,floatfix,nofootinbib,longbibliography]{revtex4-1}
\usepackage{amsmath,amssymb,color,comment,physics}
\usepackage[makeroom]{cancel}
\usepackage[caption=false]{subfig}
\usepackage{mathrsfs}
\usepackage{graphicx}
\usepackage{subfig}
\usepackage[countmax]{subfloat}
\usepackage[english]{babel}
 \usepackage[dvipsnames]{xcolor}
\usepackage[bookmarks=true,colorlinks,linkcolor=OrangeRed,urlcolor=NavyBlue,citecolor=RoyalBlue]{hyperref}
\usepackage{ulem} 
 \usepackage{dsfont}

\definecolor{mygreen}{rgb}{0,0.5,0}
\definecolor{myblue}{rgb}{0,0,0.75}
\definecolor{mymagenta}{cmyk}{0,1,0,0.12}
\definecolor{mygray}{rgb}{0.5,0.5,0.5}

\def\d{\mathrm d}

\definecolor{mypink1}{rgb}{0.858, 0.188, 0.478}
\definecolor{mypurple}{rgb}{0.49,0.18,0.56}
\definecolor{mygold}{rgb}{0.93,0.69,0.13}
\definecolor{mygreen}{rgb}{0,0.5,0}
\definecolor{myblue}{rgb}{0,0,0.75}
\definecolor{mymagenta}{cmyk}{0,1,0,0.12}
\definecolor{mygray}{rgb}{0.5,0.5,0.5}

\newcommand{\canc}[1]{}

\begin{document}

\title{Staircase Prethermalization and Constrained Dynamics in Lattice Gauge Theories}
\author{Jad C.~Halimeh}
\affiliation{INO-CNR BEC Center and Department of Physics, University of Trento, Via Sommarive 14, I-38123 Trento, Italy}
\affiliation{Kirchhoff Institute for Physics, Ruprecht-Karls-Universit\"{a}t Heidelberg, Im Neuenheimer Feld 227, 69120 Heidelberg, Germany}
\affiliation{Institute for Theoretical Physics, Ruprecht-Karls-Universit\"{a}t Heidelberg, Philosophenweg 16, 69120 Heidelberg, Germany}

\author{Philipp Hauke}
\affiliation{INO-CNR BEC Center and Department of Physics, University of Trento, Via Sommarive 14, I-38123 Trento, Italy}
\affiliation{Kirchhoff Institute for Physics, Ruprecht-Karls-Universit\"{a}t Heidelberg, Im Neuenheimer Feld 227, 69120 Heidelberg, Germany}
\affiliation{Institute for Theoretical Physics, Ruprecht-Karls-Universit\"{a}t Heidelberg, Philosophenweg 16, 69120 Heidelberg, Germany}

\begin{abstract}
The dynamics of lattice gauge theories is characterized by an abundance of local symmetry constraints. Although errors that break gauge symmetry appear naturally in NISQ-era quantum simulators, their influence on the gauge-theory dynamics is insufficiently investigated. As we show, a small gauge breaking of strength $\lambda$ induces a staircase of long-lived prethermal plateaus. The number of prethermal plateaus increases with the number of matter fields $L$, with the last plateau being reached at a timescale $\lambda^{-L/2}$, showing an intimate relation of the concomitant slowing down of dynamics with the number of local gauge constraints. Our results bode well for NISQ quantum devices, as they indicate that the proliferation timescale of gauge-invariance violation is counterintuitively delayed exponentially in system size.
\end{abstract}

\date{\today}
\maketitle
Lattice gauge theories form a powerful framework to describe the properties of fundamental particles and exotic emergent phases of matter \cite{Cheng_book,Weinberg_book,Rothe_book}. 
Despite significant results in computing static properties of gauge theories \cite{Gattringer2009}, their out-of-equilibrium dynamics remains poorly understood. 
By definition, the dynamics of a gauge theory is intrinsically governed by the gauge symmetry, which generates local constraints that have to be fulfilled at each point in space and time. 
As has been realized in recent works, this abundance of local constraints can lead to exceedingly slow dynamics, characterized by many-body scars and many-body localization-like behavior, which apparently can hinder thermalization throughout long evolution times \cite{Smith2017a,Smith2017b,Brenes2018,Turner2018,Karpov2020}. 
Despite such advances, the question of how and when these locally constrained systems thermalize is still wide open \cite{Berges2018,Busza2018}.

\begin{figure}[htp]
	\centering
	\hspace{-.25 cm}
	\includegraphics[width=.48\textwidth]{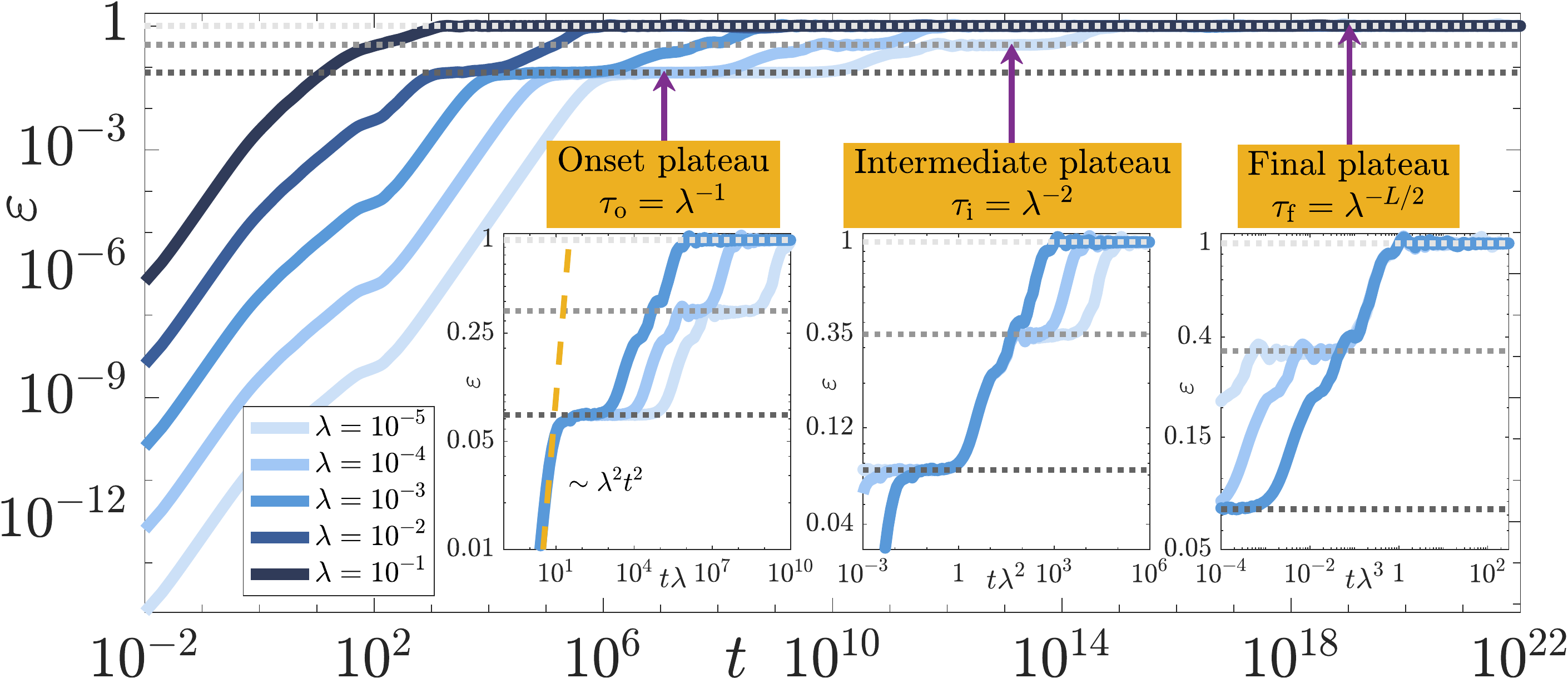}\quad
	\hspace{-.15 cm}
	\caption{(Color online). Spatiotemporal average of the gauge-invariance violation $\sum_j\langle G_j\rangle$ in the $\mathrm{Z}_2$ gauge theory with $L=6$ matter sites. As clearly seen in the insets with rescaled time axes, three distinct timescales emerge: $\tau_\mathrm{o}=1/\lambda$ for the first (onset) plateau, $\tau_\mathrm{i}=1/\lambda^2$ for the second (intermediate) plateau, and $\tau_\mathrm{f}=1/\lambda^3$ for the third ($=L/2$) and final plateau where the probability to locally find the system in either gauge-invariant manifold is $1/2$. The number of plateaus directly depends on the number of local gauge symmetries, resulting in a delay of the final plateau that scales exponentially in system size.}
	\label{fig:Fig1} 
\end{figure}

Here, we analyze the long-time dynamics of lattice gauge theories when a small perturbation explicitly breaks the gauge symmetry. 
Naively, one might expect the system to simply thermalize, as it is now described by a generic interacting many-body model without any particular local symmetry \cite{DAlessio_review,Mori_review}. 
Surprisingly, however, we find the system to undergo a series of stable prethermal plateaus, which can be well separated by many orders of magnitude of evolution time (see Fig.~\ref{fig:Fig1}). 
Our results are based on numerically exact calculations of a $\mathrm{Z}_2$ lattice gauge theory in one spatial dimension as well as analytic arguments based on a Magnus expansion \cite{Blanes2009}.  
(As we illustrate in the joint submission \cite{Halimeh2020c} for the example of a $\mathrm{U}(1)$ gauge symmetry, our findings carry over to continuous gauge groups.)  
Importantly, we find the number of plateaus to increase linearly with system size, with the last plateau being shifted to larger times, indicating that this phenomenon displays robustness against finite-size effects.

Our findings are highly relevant for quantum simulator experiments that seek to investigate gauge-theory dynamics in table-top quantum devices such as those based on trapped ions \cite{Martinez2016,Kokail2019}, superconducting qubits \cite{Klco2018,Klco2019}, or ultracold atomic gases \cite{Goerg2019,Schweizer2019,Mil2019,Yang2020}. 
Without unrealistic fine-tuning, experiments where the gauge theory emerges from microscopic processes \cite{Goerg2019,Schweizer2019,Mil2019,Yang2020} will always be plagued by residual gauge-breaking terms. 
It is currently not clear in how far such terms compromise the target gauge-theory dynamics \cite{Halimeh2020a,Halimeh2020d,Halimeh2020e}.
As our results show, the intrinsic dynamics of the system itself can halt the loss of gauge symmetry, such that the state retains a memory of its initial gauge sector for very long times. 

Moreover, the observed plateaus can be seen as a novel scenario for prethermalization due to breaking of a local symmetry. Most commonly, prethermalization appears when the integrability of a model is slightly lifted, giving rise to one long-lived plateau before the steady state is reached \cite{Mori_review}. Instead, in the present scenario of an interacting many-body system that is close to a gauge theory, an entire prethermalization landscape appears with a multitude of plateaus.

\textbf{\emph{Model and quench protocol.---}}We consider a one-dimensional $\mathrm{Z}_2$ lattice gauge theory \cite{Zohar2017,Barbiero2019,Borla2019} inspired by a recent cold-atom experiment \cite{Schweizer2019}. The model is given by the Hamiltonian
\begin{align}\label{eq:H0}
H_0=&\,\sum_{j=1}^L\big[J_a\big(a^\dagger_j\tau^z_{j,j+1}a_{j+1}+\mathrm{H.c.}\big)-J_f\tau^x_{j,j+1}\big],
\end{align} 
with $L$ matter sites and periodic boundary conditions. 
The matter fields are represented by hard-core bosons with the ladder operators $a_j,a_j^\dagger$, residing on matter sites $j=1,\ldots,L$ and obeying the canonical commutation relations $[a_j,a_l]=0$ and $[a_j,a_l^\dagger]=\delta_{j,l}(1-2a_j^\dagger a_j)$. 
The matter field on site $j$ has a local charge $Q_j=1-2a_j^\dagger a_j$. 
The electric (gauge) field on the link between matter sites $j$ and $j+1$ is represented by $\tau_{j,j+1}^{x(z)}$, which is the $x$ ($z$) component Pauli matrix. The kinetic energy of the matter field couples to the $\mathrm{Z}_2$ gauge field with a strength $J_a$, and the electric field has energy $J_f$. 

The local symmetry generators of the $\mathrm{Z}_2$ gauge group are 
\begin{align}\label{eq:Gauss}
G_j=1-(-1)^j\tau^x_{j-1,j}Q_j\tau^x_{j,j+1}, 
\end{align}
with eigenvalues $g_j=0,2$. 
Gauge invariance requires $[H_0,G_j]=0$,~$\forall j$, i.e., ideally the generators $G_j$ are conserved quantities at each matter site. In analogy to QED, this conservation is often referred to as a generalized Gauss's law.

In realistic cold-atom quantum simulators \cite{Hauke2012}, there exist microscopic processes that break gauge invariance. 
Inspired by the main error terms in the experiment of \cite{Schweizer2019}, we model these by the Hamiltonian 
\begin{align}\nonumber
H_1=\,\sum_{j=1}^L\Big[&\big(c_1a_j^\dagger\tau^+_{j,j+1} a_{j+1}+c_2a_j^\dagger \tau^-_{j,j+1} a_{j+1}+\mathrm{H.c.}\big)\\\label{eq:H1}
&\,+a_j^\dagger a_j\big(c_3\tau^z_{j,j+1}-c_4\tau^z_{j-1,j}\big)\Big],
\end{align}
where the constants $c_l$ are dependent on experimental parameters \cite{Schweizer2019}. (The salient phenomena discussed here do not depend on the precise form of $H_1$ \cite{Halimeh2020c}.) 

To mimic a typical quantum-simulator experiment, we prepare our initial state $\ket{\psi_0}$ in a product state in the gauge-invariant sector $G_j\ket{\psi_0}=0$,  $\forall j$. Specifically, we choose $2a^\dagger_ja_j\ket{\psi_0}=[1+(-1)^j]\ket{\psi_0}$ and $ \tau^x_{j,j+1}\ket{\psi_0}=(-1)^{j+1}\ket{\psi_0}$.  We then quench $\ket{\psi_0}$ with the Hamiltonian $H=H_0+\lambda H_1$, which for $\lambda\neq 0$ will drive the time-evolved state $\ket{\psi(t)}=\exp(-\mathrm{i}Ht)\ket{\psi_0}$ out of the initial gauge-invariant sector. 

\textbf{\emph{Exact diagonalization results for staircase prethermalization.---}}

As the main finding of our work, we show that during time evolution the gauge-invariance violation and local observables enter a series of prethermal plateaus at timescales $\lambda^{-s}$, with $s=0,\ldots,L/2$, with the last being the final steady-state plateau.
This finding means that full gauge noninvariance---when the wave function is equally likely to be locally in the gauge-invariant sector $g_j=0$ and $g_j=2$---sets in only at timescale $\tau_\mathrm{f}=\lambda^{-L/2}$. Counterintuitively, full gauge noninvariance is thus delayed exponentially in system size.

Towards the end of this Letter, we will provide analytic explanations based on a Magnus expansion \cite{Blanes2009,Halimeh2020c}. 
Before proceeding to the analytic arguments, however, it is instructive to illustrate this behavior using exact diagonalization calculations \cite{Johansson2012,Johansson2013,Weinberg2017,Weinberg2019}. 

We set $J_a=1$ and $J_f=0.54$ as in \cite{Schweizer2019}, but our conclusions are independent of these values \cite{Halimeh2020c}. The spatiotemporal average of the gauge-invariance violation,
\begin{align}\label{eq:error}
\varepsilon(t)=\frac{1}{Lt}\int_0^t\d \tau\sum_{j=1}^L\bra{\psi(\tau)} G_j\ket{\psi(\tau)},
\end{align}
is shown in Fig.~\ref{fig:Fig1} at several values of $\lambda$ for $L=6$ matter sites. Three ($=L/2$) distinct plateaus can be discerned, each characterized by a clear time interval during which the violation is effectively constant. 
These plateaus set in at timescales $\tau_\mathrm{o}=\lambda^{-1}$ (onset), $\tau_\mathrm{i}=\lambda^{-2}$ (intermediate), and $\tau_\mathrm{f}=\lambda^{-3}$ (final). Once the final plateau has been reached, the gauge-invariance violation is unity, indicating an equal probability of both eigenvalues $g_j=0,2$ of the gauge generator $G_j$.

In order to obtain further insight into the appearance of the prethermal plateaus, we consider projectors onto gauge-invariant supersectors, defined as the sets of gauge-invariant sectors in $H_0$ with a fixed number $s$ of nonzero gauge eigenvalues $g_j$:
\begin{align}\label{eq:P}
P_s=\sum_{\alpha_{\{s\}}}\sum_q|\alpha_{\{s\}},q\rangle\langle\alpha_{\{s\}},q|.
\end{align}
Here, $\{|\alpha_{\{s\}},q\rangle\}$ are eigenstates of $H_0$, where $\alpha_{\{s\}}$ denotes a gauge-invariant sector where $g_j\neq0$ at $s$ matter sites, and $q$ denotes all remaining good quantum numbers. Our conclusions remain the same for models with a larger number of local gauge eigenvalues, such as the $\mathrm{U}(1)$ gauge theory \cite{Halimeh2020c,Yang2016}. 

\begin{figure}[htp]
	\centering
	\hspace{-.25 cm}
	\includegraphics[width=.48\textwidth]{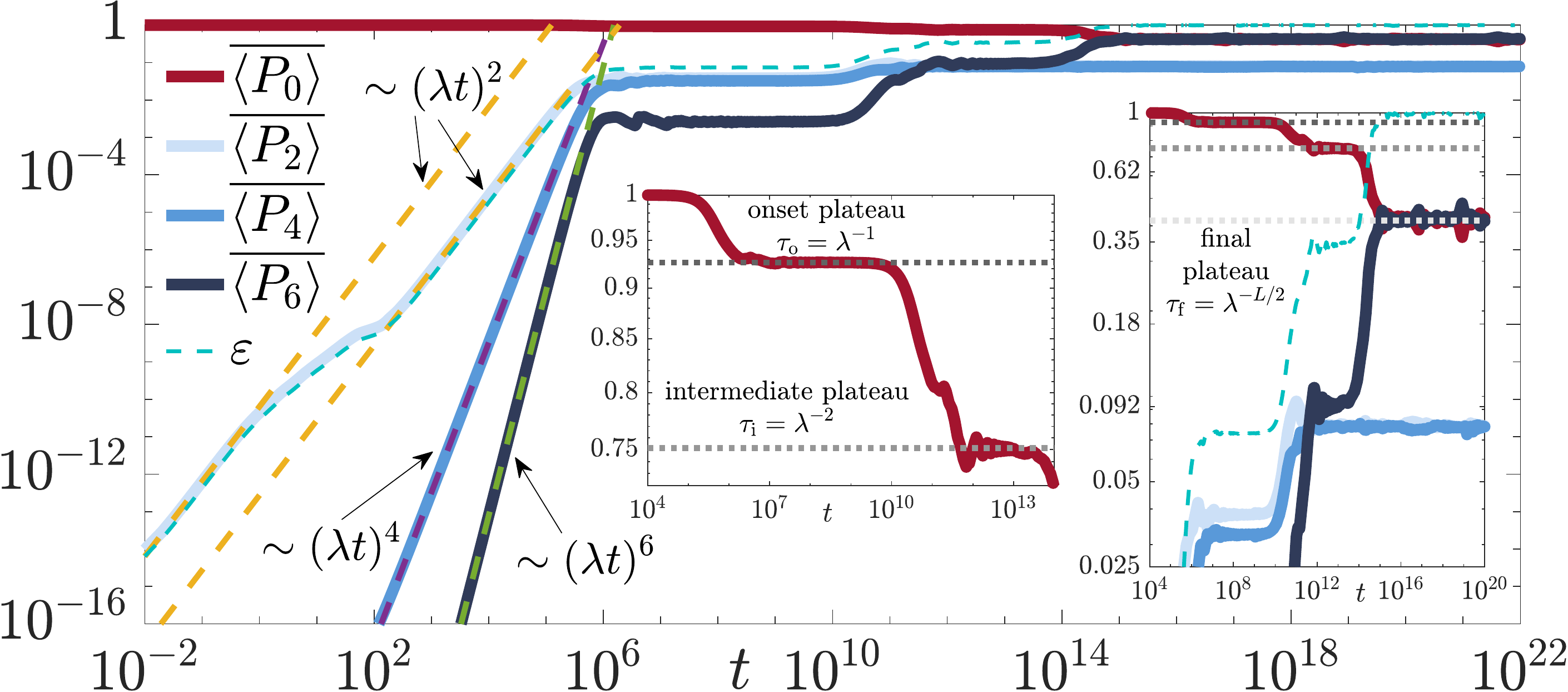}\quad
	\hspace{-.15 cm}
	\caption{(Color online). Dynamics of the time-averaged expectation values of the projectors onto gauge-invariant supersectors as defined in Eq.~\eqref{eq:P}, for $\lambda=10^{-5}$ and $L=6$ matter sites. The behavior is qualitatively the same for other values of $\lambda$. The first prethermal plateau is dominated by low-order gauge invariance-breaking processes quantified in $\langle P_2\rangle$ and $\langle P_4\rangle$. 
	In the intermediate plateau, the initial gauge-invariant supersector still dominates, while all other supersectors assume comparable values. 
	Only in the exponentially (in system size) delayed final plateau does $\langle P_6\rangle$ settle to the same value as $\langle P_0\rangle$. 
	($\langle P_s\rangle=0$ for odd $s$ because the error terms in Eq.~\eqref{eq:H1} induce simultaneous gauge-invariance breaking at an even number of local constraints.)} 
	\label{fig:Fig2} 
\end{figure}

Figure~\ref{fig:Fig2} presents the numerical expectation values $\langle P_s(t)\rangle$. (We have $\langle P_s(t)\rangle=0$ for all odd $s$, since any $H_1$ that conserves the boson number of the matter fields always breaks an even number of local Gauss's laws.)  
The projection onto the gauge-invariant sector $G_j\ket{\psi}=0,\,\forall j$, given by $\langle P_0\rangle$, remains robust around unity up to a timescale of $\lambda^{-1}$, when gauge-noninvariant processes become relevant. During this period, population in subspaces $\langle P_s\rangle$ with even $s>0$, are building up as $\sim(\lambda t)^s$. These scalings can be derived in time-dependent perturbation theory (TDPT) \cite{Halimeh2020c}. 

At timescale $\tau_\mathrm{o}=\lambda^{-1}$, $\langle P_2\rangle$ becomes large enough to drive $\langle P_0\rangle$ into a power-law decay (see insets of Fig.~\ref{fig:Fig2}) and into the onset plateau. At this moment, the violation of gauge invariance stabilizes, and $\langle P_0\rangle$ as well as $\langle P_2\rangle$, $\langle P_4\rangle$, and $\langle P_6\rangle$ reach a constant value. This plateau persists up to a timescale $\tau_\mathrm{i}=\lambda^{-2}$, where $\langle P_s\rangle$ with $s>0$ begins to grow again. $\langle P_0\rangle$ is driven into another power-law decay to an intermediate plateau where $\langle P_2\rangle$ and $\langle P_4\rangle$ settle indefinitely, with their equilibration to the same value being due to a symmetry in the spectra of $H_0$. Finally, at a timescale $\tau_\mathrm{f}=\lambda^{-3}=\lambda^{-L/2}$, processes accessing $\langle P_6\rangle$ begin to dominate and the latter grows again until it and $\langle P_0\rangle$ equilibrate to an equal value (also due to a symmetry in the energy spectra of $H_0$) in a final plateau. The latter ushers in the steady state in which both gauge eigenvalues are locally equally occupied.

\begin{figure}[htp]
	\centering
	\hspace{-.25 cm}
	\includegraphics[width=.49\textwidth]{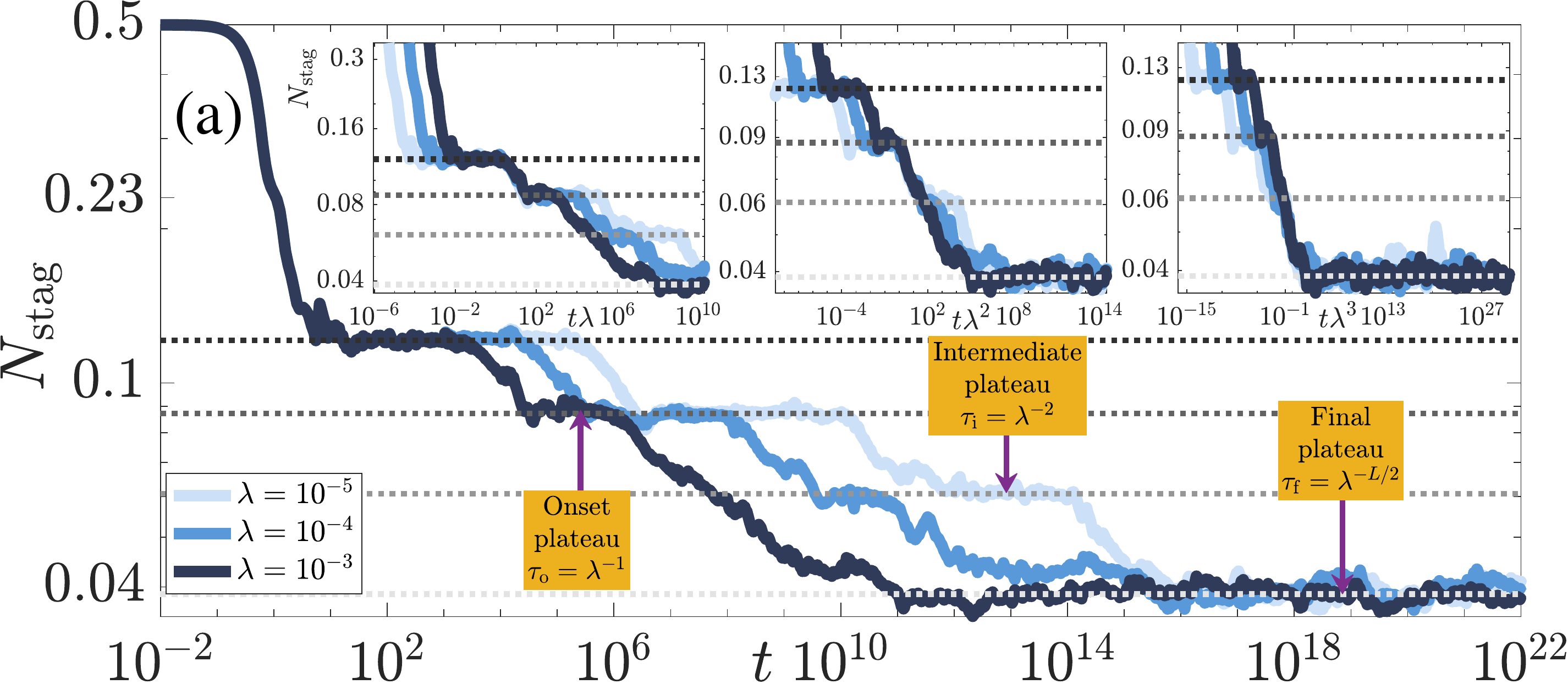}\quad\\
	\hspace{-.25 cm}
	\includegraphics[width=.49\textwidth]{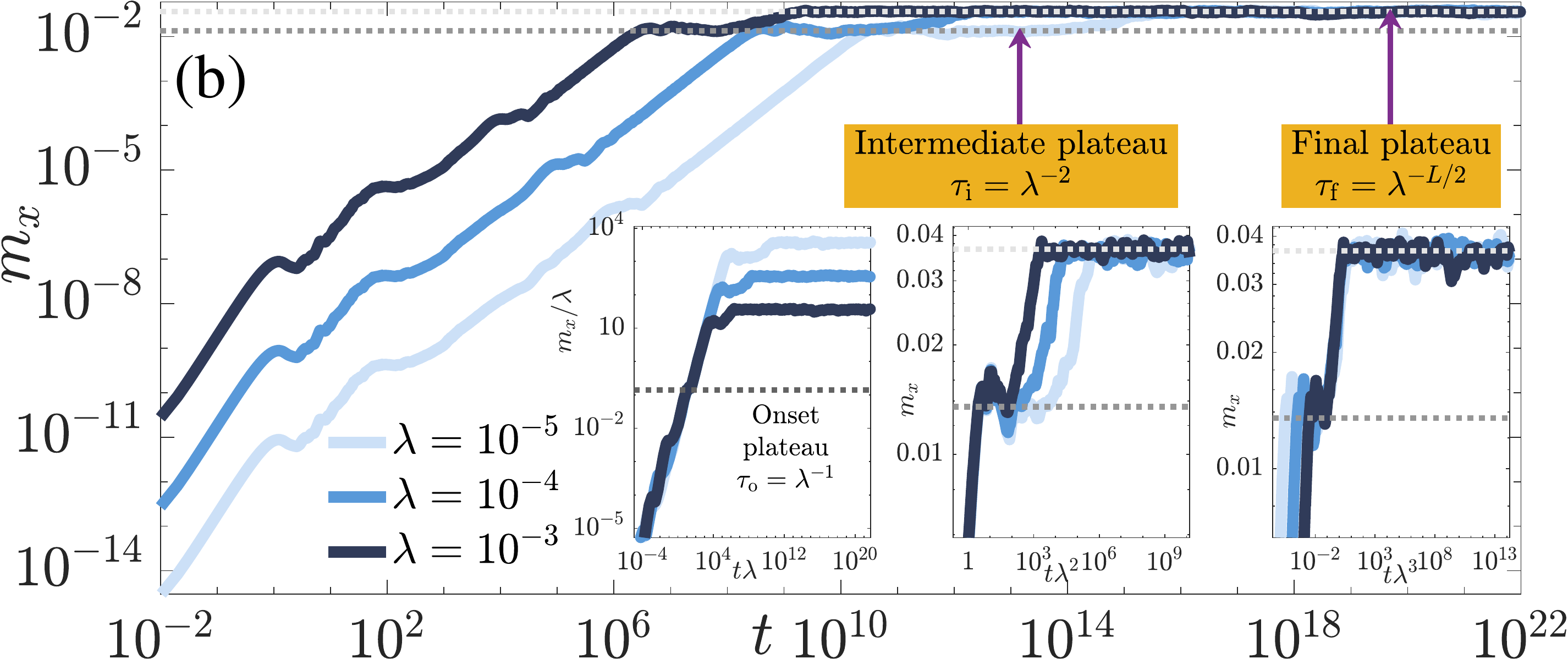}\quad
	\hspace{-.25 cm}
	\caption{(Color online).  Dynamics of the spatiotemporally averaged expectation values of (a) the staggered boson occupation and (b) the electric field. A clear staircase of three plateaus can be seen also for these local observables. (The timescales are shifted with respect to $\epsilon$ as in Fig.~\ref{fig:Fig1}, because the observables here do not commute with $H_0$.)} 
	\label{fig:obs} 
\end{figure}

This prethermalization effect also influences local observables, such as the spatiotemporal averages of the staggered boson number
\begin{align}
N_\mathrm{stag}=\frac{1}{Lt}\int_0^t\d \tau\,\Big|\sum_{j=1}^L(-1)^j\bra{\psi(\tau)}a_j^\dagger a_j\ket{\psi(\tau)}\Big|,
\end{align}
and the electric field
\begin{align}
m_x=\frac{1}{Lt}\int_0^t\d \tau\,\Big|\sum_{j=1}^L\bra{\psi(\tau)}\tau^x_{j,j+1}\ket{\psi(\tau)}\Big|.
\end{align}
As can be seen in Fig.~\ref{fig:obs} for $L=6$ matter sites, both observables show a clear three-stage plateau structure in agreement with similar behavior in the gauge-invariance violation in Fig.~\ref{fig:Fig1}. Again, plateaus occur at timescales $\lambda^{-1}$, $\lambda^{-2}$, and $\lambda^{-3}=\lambda^{-L/2}$.

\begin{figure}[htp]
	\centering
	\hspace{-.25 cm}
	\includegraphics[width=.49\textwidth]{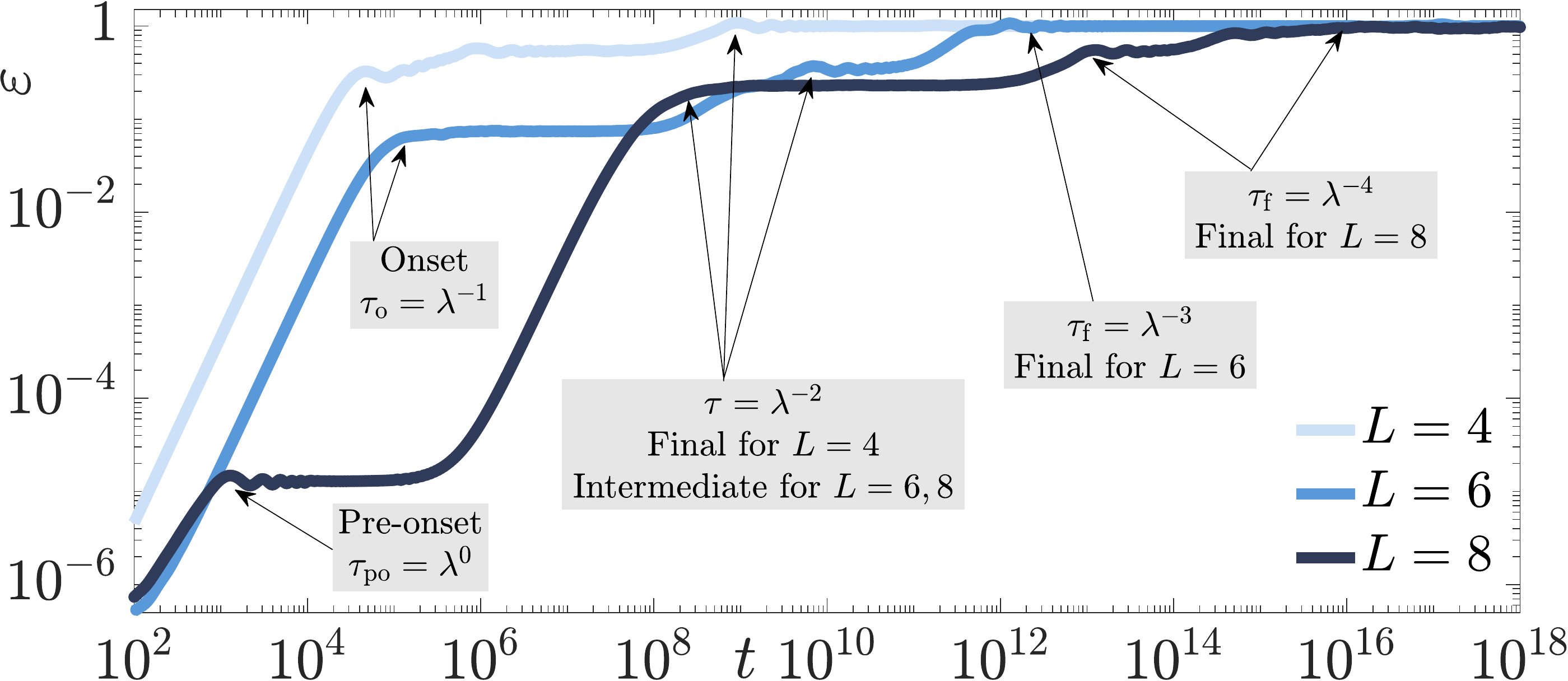}\quad
	\hspace{-.25 cm}
	\caption{(Color online). Dynamics of the spatiotemporally averaged gauge-invariance violation of Eq.~\eqref{eq:error} for $\lambda=10^{-4}$ and $L=4$, $6$, and $8$ matter sites (light to dark blue curves). The number of plateaus, including the final steady state, depends directly on the system sizes we can access in ED. Interestingly, the $s^\text{th}$ plateau appears suppressed with system size, except when $s=L/2$, where the plateau is saturated at unity, indicating full violation of gauge invariance. The timescales of the plateaus are $\lambda^{-s}$, with $s=0,\ldots,L/2$ (see text).} 
	\label{fig:Fig3}
\end{figure}

The observed \textit{staircase prethermalization} behavior is intrinsically connected to the number of matter sites $L$, or better said, to the number of local constraints due to the Gauss's law generators $G_j$. 
To illustrate this point, we consider the dynamics of the gauge-invariance violation (see Eq.~\eqref{eq:error} and Fig.~\ref{fig:Fig1}), for fixed $\lambda=10^{-4}$ and compare $L=4$, $6$, and $8$ matter sites. As shown in Fig.~\ref{fig:Fig3}, the plateaus proliferate with system size. Whereas for $L=6$ matter sites there are three plateaus, there are only \textit{two} plateaus for $L=4$, and \textit{four} plateaus for $L=8$ matter sites.

\textbf{\emph{Analytic explanation.---}}We can understand the appearance of the prethermal plateaus as well as associated timescales from analytic arguments. 
At times $t\ll \lambda^{-1}$, one can perform TDPT. It correctly predicts the initial increases of gauge-invariance violation and projector expectation values as even powers of $t\lambda$ \cite{Halimeh2020a,Halimeh2020c}. The appearance of multiple plateaus as well as the associated timescales $\lambda^s$ ($s>0$) is, however, beyond the predictive power of simple TDPT. 
Instead, we can obtain the plateaus from a Magnus expansion \cite{Blanes2009}, which amounts to resumming infinite series of terms in the time-dependent perturbative expansion. 

For small $\lambda$, it is useful to separate out the dominant gauge-invariant dynamics from the full time-evolution operator $U(t)=\mathrm{e}^{-\mathrm{i}(H_0+\lambda H_1)t}=\mathrm{e}^{-\mathrm{i}H_0t}\tilde{U}(t)$, with 
\begin{equation}
	\label{eq:Utilde}
	\tilde{U}(t)=\mathcal{T}\big\{\mathrm{e}^{-\mathrm{i}\lambda\int_0^t \d\tau H_1(\tau)}\big\}.
\end{equation}
The time-ordering prescription $\mathcal{T}$ makes $\tilde{U}(t)$ in general difficult to evaluate. 
Nevertheless, the Magnus expansion allows one to find an effective exponential form, $\tilde{U}(t)=\mathrm{e}^{\Omega(t)}$, through a skew-Hermitian operator $\Omega(t)=\sum_{n=1}^\infty\Omega_n(t)$ expanded in powers of $\lambda$. 
For the following discussion, the first two expansion terms suffice, 
\begin{subequations}
\begin{align}\label{eq:Omega1}
&\Omega_1(t)=-\mathrm{i}\lambda\int_0^t \d t_1H_1(t_1),\\\label{eq:Omega2}
&\Omega_2(t)=-\frac{\lambda^2}{2}\int_0^t \d t_1\int_0^{t_1}\d t_2 \big[H_1(t_1),H_1(t_2)\big],
\end{align}
\end{subequations}
where $H_1(t)=\mathrm{e}^{\mathrm{i}H_0t}H_1\mathrm{e}^{-\mathrm{i}H_0t}$. In the Lehmann representation with respect to $H_0$, we obtain 
\begin{align}
\Omega_1(t)=&\,-\mathrm{i}\lambda\int_0^t \d t_1\sum_{\alpha,\beta}\sum_{q,l}\mathrm{e}^{\mathrm{i}(E_{\alpha,q}-E_{\beta,l})t_1}\nonumber\\
&\,\qquad\times\bra{\alpha,q}H_1\ket{\beta,l}\ket{\alpha,q}\bra{\beta,l}.
\end{align}
Here, $\ket{\alpha,q},\ket{\beta,l}$ are again eigenstates of $H_0$, where $\alpha,\beta$ are sectors with fixed gauge eigenvalues $g_j$,
while $q,l$ denote all remaining quantum numbers within that gauge sector, e.g., energy.

At this point, it is crucial to distinguish two qualitatively different contributions. 
First, if $E_{\alpha,q}\neq E_{\beta,l}$, one obtains nonresonant terms $\Omega_1^\mathrm{nonres}(t)$ where
$\int_0^t \d t_1 \mathrm{e}^{\mathrm{i}(E_{\alpha,q}-E_{\beta,l})t_1}=\mathrm{i}[1-\mathrm{e}^{\mathrm{i}(E_{\alpha,q}-E_{\beta,l})t}]/(E_{\alpha,q}-E_{\beta,l})$.
At early times, these contributions to $\tilde{U}(t)$ reproduce the power-law increase of gauge invariance as in TDPT. At a timescale much larger than the relevant energy gaps, $t J_a\gg 1$, however, the bounded oscillations $\mathrm{e}^{\mathrm{i}(E_{\alpha,q}-E_{\beta,l})t}$ average away, leaving a constant gauge violation. This leads to the pre-onset plateau seen in Fig.~\ref{fig:Fig3} for $L=8$ as well as the feature in Fig.~\ref{fig:Fig1} at $t\approx 10^2/J_a$, with both occurring at timescale $\tau_\mathrm{po}=\lambda^{0}$. 

At this stage, the second contribution $\Omega_1^\mathrm{res}(t)$ stemming from resonant processes $E_{\alpha,q}=E_{\beta,l}$ becomes dominant. For such terms, we have $\int_0^t \d t_1 \mathrm{e}^{\mathrm{i}(E_{\alpha,q}-E_{\beta,l})t_1}=t$. Thus, $\Omega_1^\mathrm{res}(t)$ assumes the role of a time-independent Hamiltonian $H_\mathrm{eff}^{(1)}=\mathrm{i}\Omega_1^\mathrm{res}(t)/t\propto \lambda$, whose contribution to $\tilde{U}(t)$ generates a further increase of gauge violation, until the system reaches a steady state with respect to $H_\mathrm{eff}^{(1)}$. At this point, which is beyond the validity of TDPT, the system has reached the onset plateau, characterized by timescale $\tau_\mathrm{o}=\lambda^{-1}$.  

Following analogous considerations, $\Omega_2(t)$ has two qualitatively different contributions (see \cite{Halimeh2020c} for explicit formulas). First, nonresonant contributions average to a constant at times $t \gg1/J_a$, adding a subleading shift to the pre-onset plateau. 
Second, resonances generate a time-independent Hamiltonian $H_\mathrm{eff}^{(2)}=\mathrm{i}\Omega_2^\mathrm{res}(t)/t\propto \lambda^2$. Due to its parametric weakness, the dynamics generated by $H_\mathrm{eff}^{(2)}$ remains irrelevant up to a timescale $\tau_\mathrm{i} = \lambda^{-2}$. Once $\tau_\mathrm{i}$ is reached, $H_\mathrm{eff}^{(2)}$ induces further excitations into other gauge sectors, until the system again settles into a steady state, now with respect to $H_\mathrm{eff}^{(1)}+H_\mathrm{eff}^{(2)}$. The next plateau has been reached. 
The same reasoning can be repeated for all $\Omega_s(t)$, giving rise to a series of plateaus at timescales $\lambda^{-s}$, with $s\leq L/2$. With the final plateau at $\tau_\mathrm{f}= \lambda^{-L/2}$, both eigenvalues of the local gauge generator are equally likely. Importantly, the prethermal staircase has delayed this full gauge violation to exponentially long times in system size.

Our arguments rely on a separation of timescales, generated by the fact that applications of $H_1$ either connect states that are separated by large energy gaps corresponding to high-frequency oscillations, or that they access exactly degenerate states giving rise to a series of effective Hamiltonians whose strengths for small $\lambda$ differ by orders of magnitude \cite{Halimeh2020c}. We cannot assure this separation to persist in the thermodynamic limit, where typical many-body spectra become dense (and where the Magnus expansion may fail to converge). We can, however, note the importance of these results for current quantum simulations, which are concerned with rather small systems of at most several dozens of sites \cite{Kokail2019,Bernien2017,Yang2020}. 

Even more, we can see a strong difference from a similar setting where $H_0$ respects a global symmetry, which is then broken by some $H_1$. In such a case, the conservation is not of local gauge generators $G_j$, but of a total charge $\sum_j G_j$ (e.g., total particle number in a Bose--Hubbard model with a global $\mathrm{U}(1)$ symmetry). In a typical situation, the global charges are free to move through the system (in contrast to the breaking of a local gauge symmetry, where the nonzero gauge eigenvalues are localized by definition since $H_0$ commutes with all $G_j$). Thus, global charges gain kinetic energy and their spectrum spreads into a broad energy band. Instead of finding many exact degeneracies corresponding to gauge violations localized at different sites, $H_1$ can then access states that lie at arbitrary energetic distances within an energy band. These translate into a broad range of accessible frequencies, as well as the disappearance of resonant terms that generate the $H_\mathrm{eff}^{(s)}$. 
As a consequence, the separation of timescales is no longer given in case of breaking a global symmetry \cite{Halimeh2020c}. 

\textbf{\emph{Summary.---}}
Using exact diagonalization and analytic arguments based on a Magnus expansion, we have shown the existence of a multitude of plateaus in the dynamics of lattice gauge theories subjected to gauge invariance-breaking errors. As shown in \cite{Halimeh2020c}, our conclusions remain valid for various different initial states, including those that lie in gauge-invariant sectors different from $G_j\ket{\psi_0}=0$, $\forall j$, as well as for $\mathrm{U}(1)$ gauge theories. 
	
Our results lead to an intriguing and counterintuitive conclusion: The dynamics of an error-prone lattice gauge theory itself stabilizes gauge invariance and full gauge violation is delayed exponentially in system size, at least for small-to-intermediate size gauge theories. This is a very positive message to modern experimental implementations of lattice gauge theories in NISQ devices consisting of a few dozen sites. Moreover, our work paves the way for several immediate research questions. Will this behavior persist in the thermodynamic limit? Or is there a maximal size $L_\mathrm{max}$ at which the many-body energy levels become too close for the Magnus expansion to converge? And, assuming the former holds, how does this form of constrained dynamics relate to many-body localization dynamics in lattice gauge theories \cite{Brenes2018,Halimeh2020c}?

\textbf{\emph{Acknowledgements.---}}The authors acknowledge support by Provincia Autonoma di Trento, the DFG Collaborative Research Centre SFB 1225 (ISOQUANT), and the ERC Starting Grant StrEnQTh (Project-ID 804305).
\bibliography{PrethermalBiblio}
\end{document}